\documentclass{ws-procs9x6}

\newcommand{\beq}{\begin{equation}}
\newcommand{\eeq}{\end{equation}}
\newcommand{\bea}{\begin{eqnarray}}
\newcommand{\eea}{\end{eqnarray}}
\newcommand{\non}{\nonumber}
\newcommand{\dg}{\dagger}
\newcommand{\ra}{\rightarrow}
\newcommand{\g}{\gamma}
\renewcommand{\d}{\delta}

\renewcommand{\r}{\rho}
\renewcommand{\b}{\beta}

\newcommand{\m}{\mu}

\newcommand{\s}{\sigma}

\newcommand{\D}{\Delta}

\newcommand{\oh}{\frac{1}{2}}

\newcommand{\rf}[1]{(\ref{#1})}
\newcommand{\mycite}[1]{[\refcite{#1}]}

\begin{document}

\title{Adventures in Coulomb Gauge
\footnote{\uppercase{T}alk presented by \uppercase{J}. \uppercase{G}reensite
at {\sl \uppercase{C}onfinement 2003}, \uppercase{T}okyo, 
\uppercase{J}uly 21-24.}
\footnote{\uppercase{W}ork supported by the \uppercase{US D}ept.\ of
\uppercase{E}nergy, \uppercase{G}rant \uppercase{N}o.\
\uppercase{DE-FG}03-92\uppercase{ER}40711 (\uppercase{J.G.}), and
the \uppercase{S}lovak \uppercase{G}rant \uppercase{A}gency for
\uppercase{S}cience, \uppercase{G}rant \uppercase{N}o.\ 2/3106/2003
(\uppercase{\v S}.\uppercase{O}.) } }

\author{Jeff Greensite}

\address{Physics and Astronomy Dept.,
San Francisco State University, \\
San Francisco, CA 94117.
E-mail: greensit@stars.sfsu.edu}

\author{{\v S}tefan Olejn\'{I}k}

\address{Institute of Physics, Slovak Academy of Sciences, \\
 SK-845 11 Bratislava, Slovakia.
E-mail: fyziolej@savba.sk}


\maketitle

\abstracts{ We study the phase structure of SU(2) gauge theories
at zero and high temperature, with and without scalar matter
fields, in terms of the symmetric/broken realization of the remnant
gauge symmetry which exists after fixing to Coulomb gauge.  The
symmetric realization is associated with a linearly rising color
Coulomb potential (which we compute numerically), and is a necessary
but not sufficient condition for confinement.}

   There are several reasons why Coulomb gauge may be interesting
and/or useful in the study of the confining force.  First of all
there is the speculation by Gribov \mycite{Gribov} and Zwanziger
\mycite{Dan} that confinement in Coulomb gauge is due to
instantaneous (dressed) one-gluon exchange.  Secondly, the
behavior of the color Coulomb potential, defined in Coulomb gauge,
is an important element in the
gluon-chain model of QCD string formation \mycite{gchain}.  Finally,
as we will see below, the confining property of the color Coulomb
potential is associated with the unbroken realization of a remnant
gauge symmetry, and this suggests a new order parameter for studying
the phase structure of lattice gauge theories.

   We begin with the idea that confinement arises
from one-gluon exchange in Coulomb gauge; specifically, from the
instantaneous piece of the $\langle A_0 A_0 \rangle$ propagator
\beq
        \langle A^a_0(x) A^b_0(y) \rangle = P(\vec{x}-\vec{y})
             \d^{ab} \d(x_0-y_0) +  \mbox{non-instantaneous}
\eeq
where
\beq
     P(\vec{x}-\vec{y})\d^{ab} = \left\langle \left[\frac{1}{\nabla \cdot D(A)}
   (-\nabla^2) \frac{1}{\nabla \cdot D(A)} \right]_{x,y}^{a,b} \right\rangle
\eeq
where $D_i(A)$ is the covariant derivative. 
This quantity is directly related to the Coulomb interaction
energy in Coulomb gauge.

   Recall that the classical Hamiltonian in Coulomb gauge, 
$H=H_{glue}+H_{coul}$, has the form
\bea
H_{glue} &=& \oh \int d^3x ~ (E^{tr,a}\cdot E^{tr,a}  + B^a \cdot B^a)
\non \\
H_{coul} &=& \oh \int d^3x d^3y ~\r^a(x) K^{ab}(x,y;A) \r^b(y)
\non \\
K^{ab}(x,y;A) &=& \left[ \frac{1}{\nabla \cdot D(A)} (-\nabla^2)
                \frac{1}{\nabla \cdot D(A)} \right]^{ab}_{xy}
\non \\
\r^a &=& \r_{matter}^a - g f^{abc} A^b_k E^c_k
\eea
Note that $\langle K \rangle$ is the instantaneous piece of
the $\langle A_0 A_0 \rangle$ propagator.  Gribov and Zwanziger
argue that this propagator is enhanced by configurations at the
Gribov horizon, defined as a boundary in function space where the
operator $\nabla \cdot D(A)$ aquires a zero eigenvalue.  The
conjecture is that this enhancement leads to a confining Coulomb
potential, and therefore confinement by one-gluon exchange.

   One objection to this idea is that it is difficult to see how
the string-like properties of the QCD flux tube, namely, the
logarithmic growth of the flux tube cross-section (roughening), and
the universal $-\pi/12 R$ contribution to the static quark potential
(the L\"uscher term), could arise from one-gluon exchange.  On the
other hand, as we will see below, the color Coulomb
potential is an upper bound on the confining static quark potential.
This means that confinement by one-gluon exchange is a
necessary condition for confinement.

   Let
\beq
|\Psi_{qq}\rangle = \overline{q}^a(0) q^a(R) |\Psi_0 \rangle
\label{pqq}
\eeq
be a physical state in Coulomb gauge containing two static charges;
$\Psi_0$ is the ground state.  Then
\bea
         \D E &=& \langle \Psi_{qq}|H|\Psi_{qq}\rangle
                - \langle \Psi_0|H|\Psi_0\rangle
\non \\
                &=& V_{coul}(R) + E_{se}
\eea
is the expectation value of the excitation energy, where the
$R$-dependent Coulomb potential $V_{coul}(R)$ can only arise from the
expectation value of the non-local $\r K \r$ piece of the Hamiltonian.
We want to address the following questions: Is $V_{coul}(R)$ confining?
If so, is it asymptotically linear?  If it is linear, then is the
associated string tension $\s_{coul}$ equal to string tension $\s$ of the
static quark potential?  Finally, we would like to study the effect, on
the Coulomb string tension, of removing center vortices.

   We begin by defining the correlator, for SU(N) gauge theory 
in Coulomb gauge, of two timelike Wilson lines
\bea
    G(R,T) &=& \langle \frac{1}{N} \mbox{Tr}[L^\dg(0,T) L(R,T)] \rangle
\non \\
           &=& \langle \Psi_{qq} | e^{-(H- E_0)T} | \Psi_{qq} \rangle
\eea
where
\beq
L(\vec{x},T) = P\exp\left[i\int_0^T dt A(\vec{x},t)\right]
\eeq
Note that $L$ is a timelike Wilson line (\emph{not} a Polyakov line) of time extent
$T$.  The existence of a transfer matrix implies
\beq
  G(R,T) = \sum_n \left| \langle \Psi_n|\Psi_{qq}\rangle \right|^2 
                    e^{-\D E_n T}
\eeq
where the sum is over energy eigenstates, and $\D E_n$ is the energy above
the ground state.  Denote
\beq
V(R,T) = -\frac{d}{dT} \log[G(R,T)]
\eeq
Then its not hard to see that
\bea
     \D E &=& V_{coul}(R) + E_{se}
\non \\
              &=& V(R,0)
\eea
while
\bea
     \D E_{min} &=& V(R) + E'_{se}
\non \\
                    &=& \lim_{T\ra \infty} V(R,T)
\eea
where $\D E_{min}$ is the minimum energy of the $q\overline{q}$ system, and
$V(R)$ is the static quark potential.  The use of Wilson line correlators,
in Coulomb gauge, to compute the static potential was first suggested by
Marinari et al.\ \mycite{Marinari}.

  With a lattice regularization, the self-energies $E_{se}$ and
$E'_{se}$ become negligible, at large $R$, compared to the confining static
potential $V(R)$.  Then, since $\D E_{min} \le \D E$, it follows that
asymptotically
\beq
       V(R) \le V_{coul}(R)
\eeq
This inequality was originally derived by Zwanziger \mycite{Dan1}. It
implies that if confinement exists, it exists already at the level of
dressed one-gluon exchange.

   With a lattice regularization, we have
\bea
L(x,T) &=& U_0(x,a) U_0(x,2a) \cdot \cdot \cdot U_0(x,T)
\non \\
V(R,T) &=& \frac{1}{a} \log\left[\frac{G(R,T)}{G(R,T+a)} \right]
\eea
so that
\bea
\lim_{\b \ra \infty} V(R,0) &=& V_{coul}(R) ~+~ \mbox{const.}
\non \\ \non \\
\lim_{T \ra \infty} V(R,T) &=& V(R) ~+~ \mbox{const.}
\eea
where, in lattice units $a=1$,
\beq
V(R,0) = - \log[G(R,1)]
\eeq
Via lattice Monte Carlo we can then arrive at an estimate, exact in the
continuum limit, of $V_{coul}(R)$ from $V(R,0)$, and compare this
to the static quark potential $V(R)$.

\begin{figure}[ht]
\centerline{\epsfxsize=3.0in\epsfbox{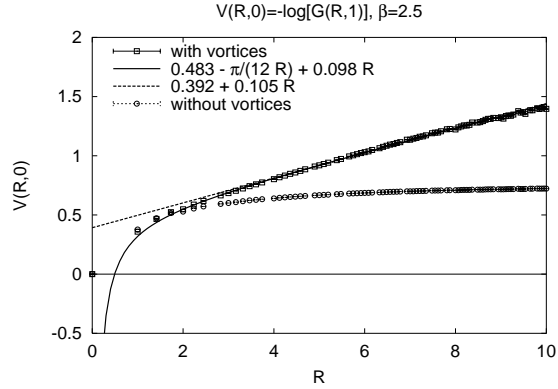}}
\caption{Lattice approximation $V(R,0)$ (upper data set, open squares)
to the color Coulomb potential $V_{coul}(R)$ + constant.  The lines are
best fits to linear, and linear + L\"uscher, potentials.  The lower data set
(open circles) for $V(R,0)$ is obtained after removing center vortices from
the lattice configurations, as described in the text.  }
\label{v0}
\end{figure}

   This procedure was carried out in ref.\ \mycite{JS} for SU(2)
lattice gauge theory.  The result for $V(R,0)$ at $\b=2.5$ is shown in
Fig.\ \ref{v0} (upper set of data points).  The two lines shown are
best fits to the data by a linear, and by a linear +
L\"uscher, potential.  The data immediately answers three of the questions
posed above: The Coulomb potential is confining, and it is linear (see also
\mycite{CZ}).
However, it turns out that the associated string tension $\s_{coul}$
is substantially greater than $\s$, by almost a factor of three.  This
means that the QCD flux tube is not simply the static charges and
their associated Coulomb (longitudinal color electric) field. The
minimum energy string state is more complicated than \rf{pqq}; it must
also contain some constituent gluons, as in the gluon chain picture of
string formation advocated in \mycite{gchain}.

  The same figure shows the effect, on $V(R,0)$, of center vortex
removal.  The center vortex theory of confinement has been studied very
actively in recent years; the theory and the numerical evidence in its
favor are reviewed in ref.\ \mycite{review}.  Center vortices are identified
by first fixing to an adjoint gauge, and then projecting link variables to
the center subgroup of the gauge group.  An example is the direct maximal
center gauge (= Landau gauge in the adjoint representation),
where the procedure is to gauge fix to a local maximum of
\beq
R = \sum_{x,\m} \Bigl| \mbox{Tr}[U_\m(x)] \Bigr|^2
\eeq
and then to project each link to the closest center element, e.g.\
for SU(2)
\beq
U_\m(x) \ra Z_\m(x) = \mbox{signTr}[U_\m(x)]
\eeq
Vortices are removed from a given lattice configuration by multiplying the
adjoint gauge-fixed configuration by the projected configuration, i.e.
\beq
        U_\m(x) \ra U'_\m(x) = Z_\m(x) U_\m(x)
\eeq
In ref.\ \mycite{dFE} it was shown that after vortex removal the string tension
vanishes, chiral symmetry breaking is eliminated, and each vortex-removed
configuration has zero topological charge.

   One can then ask what vortex removal does to the color Coulomb
potential.  In this connection, a relevant fact is that thin center 
vortices can be shown \mycite{coming} to lie on the Gribov horizon, which is
thought to play an important role in the enhancement of the Coulomb energy.
In our numerical study, the modified configuration $U'_\m(x)$ is gauge-fixed 
to Coulomb
gauge, the timelike link correlators are calculated, and $V(R,0)$ is extracted.
The result at $\b=2.5$ is shown in the lower data set (open circles) in Fig.\
\ref{v0}.  It is clear that vortex removal completely
removes the confining property of the Coulomb potential.  Further
details can be found in ref.\ \mycite{JS}.

   Of course, pure gauge theory at zero temperature is a special, albeit
very important, case.  In general gauge theories can exist in various
phases, and it is useful to characterize these in terms of the distribution
of the electric field emanating from a static isolated source.
\begin{itemize}
\item {\sl Massless Phase: } The electric field is spherically symmetric,
and falls off like $1/R^2$.  This is the case for compact $QED_4$, and
for lattice SU(N) gauge theory in $D>4$ dimensions, at weak couplings.
\item {\sl Confined Phase: } The color electric field is collimated into
a flux tube; global center symmetry is unbroken.  Examples include
SU(N) pure gauge theories at low temperature, and SU(N) gauge theories
with matter in the adjoint representation of the gauge group.
\item{\sl Screened Phases: } There is a Yukawa-like falloff of the color
electric field.  This happens in SU(N) gauge theories when the $Z_N$
center symmetry is broken spontaneously, as in high temperature gauge
theory and gauge-Higgs theories with the Higgs field in the adjoint
representation.  Gauge theories with only a trivial center symmetry
(consisting only of the identity element) are also in the screened phase;
these theories include SU(N) gauge theories with matter in the fundamental
representation, and $G_2$ gauge theory with or without matter fields.
\end{itemize}
For the purpose of studying Coulomb energy in these various phases, we find
it useful to introduce a new order parameter, related to the realization
of a remnant gauge symmetry in Coulomb gauge.

   Let $U_\m(x)$ be a lattice configuration fixed to Coulomb gauge.  Note
that the Coulomb gauge condition is preserved by the gauge transformation
\bea
U_k(x,t) &\ra& g(t) U_k(x,t) g^\dg(t)
\non \\
U_0(x,t) &\ra& g(t) U_0(x,t) g^\dg(t+1)
\eea
On any time slice, this is a global transformation, and therefore can be
spontaneously broken in the following sense:  At any fixed time $t$, in
the infinite volume limit, the average of timelike link variables $U_0(x,t)$
is non-zero in any thermalized configuration.  This means that
\bea
\lim_{R\ra \infty} G(R,1) &>& 0
\non  \\
\lim_{R\ra \infty} V(R,0) &=& \mbox{finite const.}
\non \\
      \s_{coul} &=& 0
\eea
in the broken phase.  Therefore Coulomb confinement or non-confinement can
be understood as the symmetric or spontaneously broken realization,
respectively, of the remnant gauge symmetry in Coulomb gauge.

   We now introduce the order parameter $Q$, as the modulus of the spatial
average of timelike links, i.e.
\bea
U^{av}_0(t) &=& \frac{1}{L^3} \sum_{\vec{x}} U_0(\vec{x},t)
\non \\
Q &=& \left\langle \sqrt{\mbox{Tr}[U_0^{av}(t) U_0^{av \dg}(t)]}
\right\rangle
\eea
On general grounds
\beq
Q = c + \frac{b}{L^{3/2}} ~~\mbox{with~~} \left\{ \begin{array}{cl}
    c=0 & \mbox{symmetric phase} \cr
    c>0 & \mbox{broken phase} \end{array} \right.
\eeq
Thus $Q>0$ in the infinite volume limit implies that $V_{coul}(R)$ is
non-confining.  Since $V_{coul}(R)$ is an upper bound on $V(R)$, this
implies that $Q=0$ is a necessary (but not sufficient) condition for
confinement.

\begin{figure}[ht]
\centerline{\epsfxsize=3.0in\epsfbox{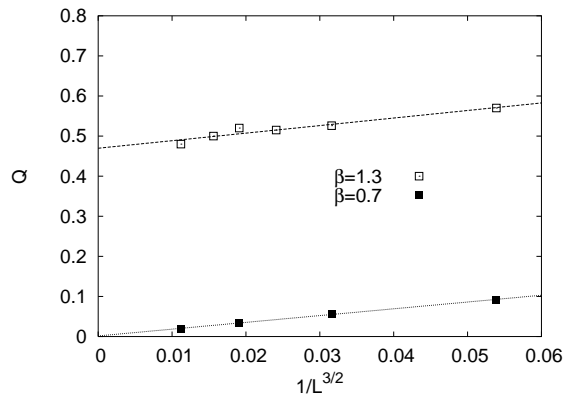}}
\caption{Extrapolation of $Q$ to infinite volume in $QED_4$, for
$\b=0.7$ (confining phase) and $\b=1.3$ (massless phase).}
\label{qv}
\end{figure}

   It is useful to try out this order parameter in compact $QED_4$, where
there is a transition from the confining to the massless phase at
$\b \approx 1$.  Figure \ref{qv} shows our results for $Q$ vs.\ the root
inverse 3-volume $L^{-3/2}$  at $\b=0.7$ (confining phase)
and $\b=1.3$ (massless phase).  In this case the $Q$ parameter seems
to nicely distinguish between the two phases, extrapolating
to zero only in the confined phase.

\begin{figure}[ht]
\centerline{\epsfxsize=3.0in\epsfbox{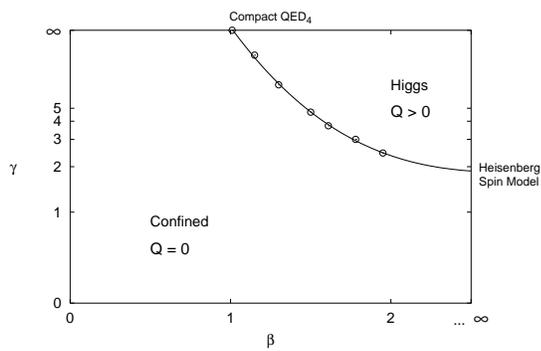}}
\caption{Phase diagram of the SU(2) adjoint Higgs model.  The plaquette
energy and the remnant symmetry order parameter $Q$ locate the same transition
line between the confined and Higgs phases.}
\label{adjhiggs}
\end{figure}

   Next, we consider SU(2) gauge-Higgs theory with a ``frozen" Higgs field in
the adjoint representation.  The lattice Lagrangian is
\bea
   S &=& \b \sum_{plaq} \oh \mbox{Tr}[UUU^\dg U^\dg]
\non \\
 &+& \frac{\gamma}{4} \sum_{x,\m} \phi^a(x)\phi^b(x+\widehat{\m})
         \mbox{Tr}[\s^a U_\m(x) \s^b U_\m^\dg(x)]
\eea
where $\phi$ is a real 3-component field satisfying the constraint
$\sum_a (\phi^a)^2 = 1$.  This is a theory with a confining, center
symmetric phase, and a non-confining phase with spontaneously broken
center symmetry.  Our finding is that the transition line in the
$\b-\g$ phase diagram corresponding to the remnant symmetry-breaking
transition is identical to the transition line for
confinement-deconfinement, mapped out long ago by Brower et al.\
\mycite{Brower} from measurements of the plaquette energy.  In the confined phase
we find $Q=0$ (when extrapolated
to infinite volume), and $Q>0$ in the Higgs phase, as indicated schematically
in Figure \ref{adjhiggs}.

\begin{figure}[t!]
\centerline{\epsfxsize=3.0in\epsfbox{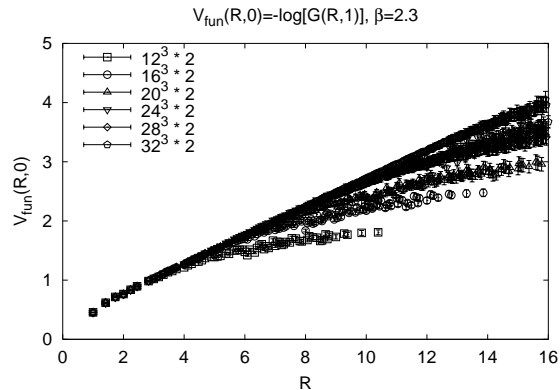}}
\caption{Coulomb potential in the deconfined phase, at $\b=2.3$ and
$L_t=2$ lattice spacings in the time direction, with spatial volumes
from $12^3$ to $32^3$.}
\label{v_deconf}
\end{figure}

\begin{figure}[t!]
\centerline{\epsfxsize=3.0in\epsfbox{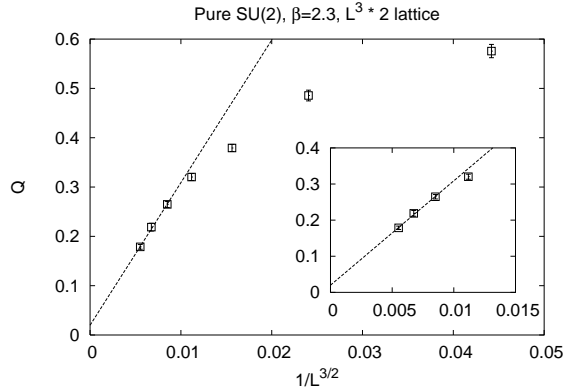}}
\caption{The $Q$ parameter vs.\ root inverse 3-volume in the deconfined
phase, $\b=2.3$ and $L_t=2$ lattice spacings.}
\label{q_deconf}
\end{figure}

   One might guess that the transition from the confined to the
deconfined phase is always accompanied by remnant symmetry breaking.
Surprisingly, this turns out not to be true.  We have also computed
$V(R,0)$ and $Q$ in the deconfined phase of pure SU(2) lattice gauge
theory, with the results shown in Figs.\ \ref{v_deconf} and 
\ref{q_deconf}.  This data was
taken at $\b=2.3$ on lattices with time extension of two lattice
spacings, well within the deconfined phase.  Yet the Coulomb potential
is clearly linear and confining at large lattice volume, 
while the extrapolation of $Q$ to infinite
volume seems compatible with zero.  A possible reason for this
behavior is the fact that $K(x,y;A)$, whose expectation value gives
the instantaneous Coulomb propagator, depends only on the spacelike
components $A_k$ at a fixed time.  On the lattice, this translates to
dependence only on spacelike links on a time slice.  But we know that
spacelike links on a time slice are a confining ensemble even in the
deconfined phase, since spacelike Wilson loops are known to have an
area law falloff at any temperature.  If the Coulomb propagator
depends only on the confining properties of spacelike links, then it
is not so surprising that the Coulomb potential is confining in the
deconfined regime (nor is this a paradox: the Coulomb potential is
only an upper limit on the static potential).  A test of this
explanation is to remove the confining properties of the spacelike
links by removing center vortices, via the de Forcrand/D'Elia
procedure explained above.  Then one expects the Coulomb potential to
be non-confining, and this is, in fact, what is observed.

  Finally, we study a gauge-Higgs system with the radially frozen
Higgs field in the fundamental representation.  For the SU(2) gauge
group, the lattice Lagrangian can be expressed as \mycite{Lang}
\bea
   S &=& \b \sum_{plaq} \oh \mbox{Tr}[UUU^\dg U^\dg]
\non \\
 &+& \gamma \sum_{x,\m} \oh
         \mbox{Tr}[\phi^\dg(x) U_\m(x) \phi(x+\widehat{\m})]
\eea
with $\phi$ an SU(2) group-valued field.  This is a theory with only
a screened phase; it can be proven that no transition to a
confined phase is possible \mycite{FS}.
There is a first-order phase transition line in the $\b-\g$ phase diagram,
but this line has an endpoint, and does not divide the
diagram into thermodynamically separate phases.

\begin{figure}[t!]
\centerline{\epsfxsize=2.5in\epsfbox{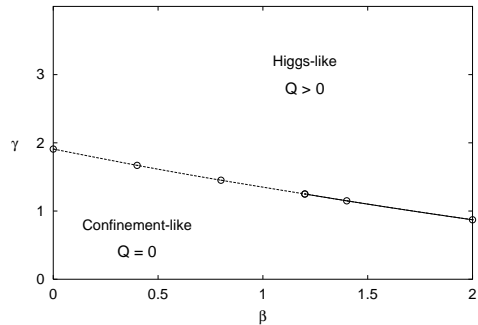}}
\caption{Phase diagram of the fundamental Higgs model.  There is a
thermodynamic transition and a $Q$ transition along the solid line,
but a non-thermodynamic transition (Kert\'esz line) in $Q$ along the
dashed line.}
\label{phase}
\end{figure}

\begin{figure}[t!]
\centerline{\epsfxsize=2.5in\epsfbox{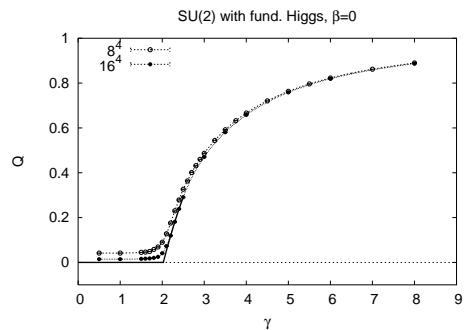}}
\caption{$Q$ vs.\ $\g$ at $\b=0$ in the SU(2) fundamental
Higgs model, on $8^4$ and $16^4$ 
lattices.  The solid line is the presumed extrapolation to infinite
volume.}
\label{q0p0}
\end{figure}

    The remnant symmetry transition line coincides with the (thermodynamic)
line of first-order transitions
found by Lang et al.\ \mycite{Lang}, but it then extends beyond the 
thermodynamic line
all the way to $\b=0$ and $\g=2$.  This line divides the phase diagram
into $Q=0$ and $Q>0$ regions, as indicated schematically in Fig.\ \ref{phase}.
In Fig.\ \ref{q0p0} we plot $Q$ vs.\ $\g$ at
$\b=0$.  If $Q$ were the magnetization of an Ising spin system, this would
surely be a second order phase transition, with the solid line in
the figure representing the infinite volume limit.  Nevertheless, there
is no thermodynamic transition.  At $\b=0$ one can easily compute
the free energy exactly, which is found to be
\beq
   F(\g) = 4V \log\left[\frac{2I_1(\g)}{\g}\right]
\eeq
This expression is perfectly analytic at all $\g>0$.  On the other hand,
a strong-coupling analysis of $G(R,1)$ at fixed $\b \ll 1$ \mycite{coming} 
arrives at an exponential decay to zero as $R\ra \infty$ at small $\g$, 
but a non-zero large-$R$ limit at large $\g$.  This implies a 
symmetry-breaking transition at some critical value $\g_{cr}(\b)$, which
motivated our numerical study of $Q$ in this model.

   The remnant symmetry breaking transition in the gauge-Higgs
system, in the absence of a thermodynamic transition, is probably an example
of a {\it Kert\'esz line} \mycite{K} in statistical mechanics.  This possibility
was first suggested by Langfeld \mycite{Kurt}, who discovered remnant gauge
symmetry
breaking in Landau gauge in a closely related model 
(see also Satz \mycite{Satz}). 
Kert\'esz lines are associated with
percolation transitions, and it is natural to ask, in this case, what is
percolating.  Recent investigations \mycite{Roman} indicate that in the
gauge-Higgs system, the Kert\'esz line locates center vortex percolation
transitions.

   This concludes a summary of our investigations of phase structure
in lattice gauge theory, as seen by Coulomb energy and remnant symmetry.
Our study has uncovered Coulomb ``over-confinement'' ($\s_{coul} \approx 3\s$)
in the low temperature confined phase,
the persistence of Coulomb confinement in the deconfined phase,
connections between vortex and Coulomb confinement, and (in accord
with Langfeld \mycite{Kurt}) symmetry
breaking in the absence of a thermodynamic phase transition.  These aspects
of non-perturbative gauge theory are somewhat surprising, and merit further
study.

\section*{Acknowledgments}

   Results reported in the second half of this talk were obtained in 
collaboration with Daniel Zwanziger.  These results will be presented in 
more detail in a later publication \mycite{coming}.  

   J.G. thanks Hideo Suganuma for the opportunity to speak 
at this enjoyable and stimulating meeting.

\end{document}